\documentclass[prb,aps,eqsecnum,twocolumn,floats,nofootinbib,superscriptaddress,showpacs]{revtex4}

\usepackage{amsmath,amssymb,dsfont,graphicx,psfrag}

\newcommand{\non}{\nonumber}
\newcommand{\nn}{\nonumber}
\newcommand{\be}{\begin{equation}}
\newcommand{\ee}{\end{equation}}
\newcommand{\bea}{\begin{eqnarray}}
\newcommand{\eea}{\end{eqnarray}}
\newcommand{\bse}{\begin{subequations}}
\newcommand{\ese}{\end{subequations}}
\newcommand{\la}{\left<}
\newcommand{\ra}{\right>}
\newcommand{\lb}{\left[}
\newcommand{\rb}{\right]}
\newcommand{\lp}{\left(}
\newcommand{\rp}{\right)}
\newcommand{\lf}{\left\{}
\newcommand{\rf}{\right\}}

\newcommand{\w}{{\omega}}
\newcommand{\W}{{\Omega}}

\newcommand{\K}{{\cal K}}

\newcommand{\tr}{{\rm tr}}

\newcommand{\sign}{\,{\rm sign}\,}

\newcommand{\A}{{\cal A}}

\newcommand{\G}{{\cal G}}

\newcommand{\ts}[1]{\textstyle{#1}}
\newcommand{\ds}[1]{\displaystyle{#1}}
\newcommand{\half}{\textstyle{\frac12}}
\newcommand{\what}{\widehat}
\newcommand{\pdag}{{\phantom{\dag}}}
\newcommand{\PRB}{Phys. Rev. B }
\newcommand{\PRL}{Phys. Rev. Lett. }

\begin{document}
\title{Critical conductance of a one-dimensional doped Mott insulator}

\author{M. Garst} 
\affiliation{Institut f\"ur
  Theoretische Physik, Universit\"at zu K\"oln, 50938 K\"oln, Germany}

\author{D. S. Novikov}
\affiliation{W. I. Fine Theoretical Physics Institute, University of
  Minnesota, Minneapolis, MN 55455, USA}

\author{Ady Stern} 
\affiliation{Department of Condensed Matter Physics, Weizmann
  Institute of Science, Rehovot 76100, Israel} 

\author{L. I. Glazman}
\affiliation{W. I. Fine Theoretical Physics Institute, University of
  Minnesota, Minneapolis, MN 55455, USA}

\date{January 22, 2008}

\begin{abstract}  
  We consider the two-terminal conductance of a one-dimensional Mott
  insulator undergoing the 
  commensurate-incommensurate quantum phase transition to
  a conducting state. We treat the leads as Luttinger liquids. At
  a specific value of compressibility of the leads, corresponding to the
  Luther-Emery point, the conductance can be described in terms of the free
  propagation of non-interacting fermions with charge $e/\sqrt{2}$.  At that
  point, the temperature dependence of the conductance across the quantum
  phase transition is described by a Fermi function. The deviation from
  the Luther-Emery point in the leads changes the temperature dependence
  qualitatively. In the metallic state, the low-temperature
  conductance is determined by the properties of the leads, and
  is described by the conventional Luttinger-liquid theory. In the
  insulating state, conductance occurs via activation of
  $e/\sqrt{2}$ charges, and is independent of the Luttinger-liquid
  compressibility.
\end{abstract}
\maketitle

\section{Introduction and Outline}

The dc conductance of a one-dimensional (1D) system crucially depends on
the properties of the leads attached to it. Unlike in higher dimensions,
this dependence is not merely a technological, but rather a generic physics problem.

In the case of the conventional Luttinger-liquid model for a quantum wire,\cite{FG}
the role of the leads in the zero-temperature ($T=0$) conductance is
well studied.\cite{KF,Safi95,Maslov95,Ponomarenko95,FG} As long as
the electrons do not backscatter off inhomogeneities, the conductance
is not sensitive to the nature of carriers in the wire, and is
determined solely by the leads. In the simplest 
case of the noninteracting 
leads, the conductance is quantized in units of
$e^2/2\pi\hbar$ per channel (each of the two spin polarizations counts
for a separate channel).  This conclusion also remains true for the
incommensurate phase of a so-called Mott wire
(doped 1D Mott insulator).\cite{Mori97,Starykh98} Thus the charge
fractionalization characteristic of these
systems~\cite{FG,GiamarchiBook} is
not observable in the zero-temperature ballistic conductance;
if the leads are Fermi liquids, then the ballistic conductance of the
wire appears to coincide with that of the free
electrons.\cite{Tarucha95}

The Tomonaga-Luttinger model is applicable to a relatively dense
system of particles forming a compressible liquid. The ``distance''
from the incompressible state may be characterized by a parameter
$-r$, having the meaning of chemical potential measured from the point
where the system becomes incompressible. For example, for free fermions
filling up a conduction band
%
\be \label{-r}
-r \equiv \mu - \Delta
\ee
with $\mu$ and $\Delta$ being the Fermi energy and band edge,
respectively.  At $T=0$ and $\mu < \Delta$ the system is incompressible.

The conventional Luttinger liquid model requires relatively
high carrier density, $-r/T\gg 1$.  How robust is the apparent
free-electron picture of transport if the condition $-r/T\gg 1$ is
violated?  One may expect deviations from the apparent free-fermion
form of the two-terminal conductance in some intermediate region of
parameters\cite{matveev} in a conventional quantum wire.  However, for
fully spin-polarized electrons (equivalent to spinless fermions) at
the point $r=0$ of the quantum phase transition the interaction
becomes irrelevant. The corresponding conductance equals
$\frac12 \times e^2/h$, i.e. a half of the unit conductance quantum $e^2/h$
for spinless fermions, 
which is again indistingushable from a free-fermion result.\cite{BHreview}   

Charge carriers at the commensurate-incommensurate quantum phase
transition in a Mott wire can also be mapped 
onto free spinless fermions.\cite{Mori97}  The distance to the phase transition is
characterized by a chemical potential $-r$, see Eq.~(\ref{-r}), here
measured with respect to the Mott gap, $\Delta$. However, these
effective fermions
have fractional charge, and are related to the 
original electrons in a complicated way. It is therefore 
an important unresolved question as to what the two-terminal critical
conductance of a Mott wire is. 

The aim of this work is to investigate
the finite-temperature conductance around the quantum phase transition
point in a Mott wire.  Our main finding is that such a setup may
provide the way to experimentally access the elusive character of the
fractional quasiparticle charge in an interacting 1D system.
 
To be specific, we consider the two-terminal conductance of a
one-dimensional electron system close to the Mott transition. The
details of how the underlying commensuration is realized in practice,
e.g.~by a periodic potential, will not be important; the only
parameter from the Mott wire we will need in the end is the effective
value $\Delta$ of the charge gap.  Next, we assume that this wire is
smoothly attached to uniform identical Luttinger liquid leads
characterized by the Luttinger parameter $K_L$. We evaluate the
finite-temperature conductance close to the Mott transition
perturbatively in $K_L-\frac{1}{2}$. The value of $K_L=\frac12$ is special
(Luther-Emery point\cite{Luther74}), as it allows a mapping of the
entire system onto a free-fermion one. In lowest order in $K_L -
\frac{1}{2}$, we find
\begin{align} \label{FinalResult}
{G\over G_0} = {\frac{1}{2}} f(r/T) + \left(K_L-\frac{1}{2}\right) f^2(r/T)\,,
\quad G_0 = \ds{2e^2\over h},
\end{align}
where $r$ is the tuning parameter (\ref{-r}) of the Mott transition, $G_0$
is the conductance quantum, and
\begin{align} \label{FermiFunction}
f(x) = \frac{1}{1+e^x}
\end{align}
is the Fermi function.
On the metallic side, $r<0$, the conductance approaches the value $K_LG_0$ 
in the limit of zero temperature, i.e., it is determined by the
Luttinger liquid parameter of the leads in agreement with
Ref.~\onlinecite{Starykh98}.  
On the other hand, on the insulating side of the transition, $r>0$,
transport at low $T$ is thermally activated as the Fermi function
reduces to a Boltzmann factor, $f(r/T) \approx e^{-r/T}$. 
In lowest order in the fugacity $e^{-r/T}$, we find that the 
conductance $G = \frac12 G_0 e^{-r/T}$
is not affected by the leads, i.e., is 
independent of $K_L$. In particular, the prefactor
of the exponential is determined by the effective conductance quantum
$\half G_0$
attributed to the degrees of freedom of the critical Mott wire. As we
will explain in detail below, the critical degrees of freedom are the fermions that carry fractional charge
\begin{align} \label{FracCharge}
e_{\rm CI} = {e}/{\sqrt{2}} \,.
\end{align}
At low temperatures transport in the Mott insulator occurs via
activation of the fractional charges (\ref{FracCharge}), resulting
in the announced prefactor $\half G_0 \equiv 2e^2_{\rm CI}/h$. 
Finally, in the extended quantum critical regime,
$|r| \ll T$, the conductance is dominated by its critical value,
\be \label{Gcr}
G_{\rm cr} \equiv \left.G\right|_{r=0} = \frac{2 K_L+1}{8} G_0 +
\mathcal{O}(K_L-\half)^2 . 
\ee
Note that the specific fraction of the
conductance quantum here depends both on the fractional charge of the
carriers and on the properties of the leads. 
These results differ from the considerations of Mori {\it et~al.}. \cite{Mori97}

Our study of the two-terminal conductance of the critical Mott insulator complements
the theoretical investigation of ballistic transport in massive
theories in general, for a recent review see
Ref.~\onlinecite{Zotos05}. The thermally activated quasiparticles
above a gapped ground state in a homogeneous system (no leads) may
travel ballistically thus giving rise to a singular
contribution to the optical conductivity, $\sigma_{\rm sing.}(\omega)
= 2 \pi D \delta(\omega)$, with a characteristic Drude weight $D =
D(T)$.
The low-temperature Drude weight of the Mott insulator is believed to
remain finite even away from criticality.\cite{Garst01} (In the exact
commensurate case of an undoped Mott insulator the fate of the Drude
weight is less clear and remains a subject of ongoing research.
\cite{Damle05,Altshuler06}) Remarkably, the knowledge of the Drude weight
in the conductivity of a uniform 1D system is
insufficient for finding its two-terminal dc conductance. The latter is affected by the different nature of the charge carriers in the leads and in the wire.  

The paper is organized as follows. In Section~\ref{sec:uniform}
we consider the properties of the uniform Mott wire. 
In Section \ref{sec:InhomSystem} we consider the Mott wire with the leads,
and evaluate the modification of the critical conductance 
induced by the presence of the leads. 
Section~\ref{sec:Summary} is devoted to the discussion of our results.

\section{Uniform Mott wire}
\label{sec:uniform}

In this Section we consider the model of the uniform 1D Mott wire.
In Sec.~\ref{sec:Model} we
introduce the Hamiltonian of our system, and in Sec.~\ref{sec:CriticalTheory} 
we sketch 
the critical theory describing the Mott transition. In Sec.~\ref{sec:Transport} 
we outline the application of the Kubo formula for
the calculation of conductance, and derive 
the critical conductance for the 
Mott wire without leads in Sec.~\ref{sec:HomSystem}.

\subsection{Model of the one-dimensional Mott insulator}
\label{sec:Model}

Consider an interacting one-dimensional electron system in the vicinity of 
the Mott transition. The latter can be realized, e.g., in a 
Hubbard model
close to half-filling. 
The effective Hamiltonian is conventionally represented in terms of the bosonic charge and spin excitations that decouple at sufficiently low energies.\cite{GiamarchiBook}
In this work we only focus on the 
charge sector which, near the transition, 
is described by the sine-Gordon model\cite{GiamarchiBook}
\begin{align}\label{SineGordon}
\mathcal{H}[\phi_c,\theta_c] = \int \frac{dx}{2\pi} \left[
K u (\partial_x \theta_c)^2 + \frac{u}{K} (\partial_x \phi_c)^2  \right]
\\\nn
+
\int dx \frac{2 V_{\rm Umkl}}{(2 \pi a)^2} \cos( \sqrt{8} \phi_c - q_0 x) \,.
\end{align}
Here $K$ is the Luttinger parameter, $u$ is the plasmon velocity, 
$V_{\rm Umkl}$ is the Umklapp scattering amplitude 
(given by the Fourier component of the lattice potential at the reciprocal 
lattice vector) that corresponds to $g_3$ 
in the standard $g$-ology classification,\cite{GiamarchiBook} 
and $a$ is a short distance cutoff.  
The conjugate fields $\phi_c$ and $\theta_c$, 
with $[\frac1\pi \partial_x \phi_c(x), \, \theta_c(x')]=-i\delta(x-x')$,
describe the smooth components of the 
density and current fluctuations within the system, 
\begin{align}
\delta n = - \sqrt{2}\, \frac{\partial_x \phi_c}{\pi} ,\quad
{\rm and}\quad
j = \sqrt{2}\, K u \frac{\partial_x \theta_c}{\pi} ,
\end{align}
respectively. The prefactor $\sqrt{2}$ in these expressions is attributed to the two spin polarizations of the electrons. 
The parameter $q_0$ measures the distance to half-filling. In the presence of repulsive interaction between the electons, $K<1$, the umklapp scattering process, $V_{\rm Umkl}$, becomes relevant and opens up a gap in the spectrum at half filling $q_0 = 0$, rendering the system into the Mott insulating phase. 

The elementary excitations of the Mott insulator are massive quantum solitons (kinks) of the model (\ref{SineGordon}). 
Classically, the kink must connect two discrete values of $\phi_c(x\to\pm\infty)$, which realize the minimum of the cosine term. This dictates the {\it fractional} 
value of the topological charge
%
\begin{align} \label{TopologicalCharge}
Q = \frac{1}{\pi} \int_{-\infty}^\infty dx\, \partial_x  \phi_c^{\rm soliton}(x) = \frac{1}{\sqrt{2}} \,.
\end{align}
Below we will absorb the topological charge (\ref{TopologicalCharge})
carried by the solitons into the fractional electric charge quantum, $e_{\rm CI} = Q e$, cf. Eq.~(\ref{FracCharge}). 
 
Upon detuning the system from half-filling by increasing $q_0$ across a critical value the system becomes conducting via the commensurate-incommensurate (CI) transition.\cite{GiamarchiBook,Giamarchi91} The ground state of this conducting phase is characterized by a finite topological charge density. Consequently, this transition is described\cite{Japaridze78,Pokrovsky79} by exploiting the duality \cite{Coleman,Haldane} between the sine-Gordon model (\ref{SineGordon}) and the massive Thirring model. To establish this duality, we make a canonical transformation
to the rescaled fields 
\be \label{phi-theta}
\phi = \sqrt{2} \phi_c\,, \quad  \mbox{and}\quad  \theta = \theta_c/\sqrt{2}.
\ee
In these variables the model (\ref{SineGordon}) becomes
\begin{align}
\mathcal{H}[\phi,\theta] = \int \frac{dx}{2\pi} \left[
2 K u (\partial_x \theta)^2 + \frac{u}{2 K} (\partial_x \phi)^2  \right]
\\\nn
+
\int dx \frac{2 V_{\rm Umkl}}{(2 \pi a)^2} \cos( 2 \phi - q_0 x) \,.
\end{align}
The above Hamiltonian for the charge sector can now be refermionized 
with the help of the standard bosonization formula,
\begin{align} \label{referm}
\psi_\kappa = \frac{1}{\sqrt{2\pi a}} e^{-i \kappa \phi + i \theta},
\end{align}
with $\kappa = +1$ and $-1$ for right (R) and left (L) movers, respectively.
The fermionic particles associated with the operator $\psi_\kappa$ are identified with the solitonic excitations of the sine-Gordon theory; the fermionic density thereby coincides with their topological charge density. The resulting fermionic theory is the massive Thirring model,
\begin{align} \label{MT}
\mathcal{H}[\Psi] &= \int dx\, \lf \vphantom{\lp\Psi^2\rp^2}
\Psi^\dag \left[
 - i v \sigma^3 \partial_x  - \mu + \sigma^1 \Delta 
\right]
\Psi \right.
\\\nn& \left.
+ \ts{\frac14} \lp {g_4 + g_2}\rp \lp\Psi^\dag \Psi\rp^2
+ \ts{\frac14}\lp g_4 - g_2\rp \lp \Psi^\dag \sigma^3 \Psi \rp^2 \rf
\end{align}
where $\sigma^i$ are the Pauli matrices, and 
we used the spinor notation $\Psi^\dag = (\psi^\dag_R, \psi^\dag_L)$
for the new {\it spinless} fermions.
As the Fermi momentum for the fictitious Dirac fermions $\Psi$ is set to 
zero (at the Dirac point), there are no spatially oscillating factors 
in Eq.~(\ref{referm}). These fermions (or the solitonic excitations of the 
sine-Gordon theory) describe the density excitations smooth
on the scale of the original lattice responsible for the Mott phase.
The parameters of the model (\ref{MT}) are the chemical potential $\mu = - v q_0/2$ 
and the gap $\Delta = V_{\rm Umkl}/(2 \pi a)$;\cite{footnote-gap} 
the velocity $v$ and the interaction constants $g_4$ and $g_2$ 
are implicitly given by 
\begin{align} \label{EffLuttingerParameters1}
u &= v \sqrt{\left(1+\nu g_4\right)^2 - \left(\nu g_2\right)^2} \,,
\\ \label{EffLuttingerParameters2}
2 K &= \sqrt{\frac{1 +\nu g_4 - \nu g_2}{1 + \nu g_4 + \nu g_2}}\,,
\end{align}
where the density of states for the $\Psi$-fermions
\be \label{nu}
\nu = {1\over 2\pi v} \,. 
\ee
Note the redundancy of the $g_4$ interaction as it enters the mapping only in combination with the velocity, $2\pi v + g_4$. We deliberately choose to keep both the
$g_2$ and $g_4$ interaction constants, as it will help us later to classify
different perturbative contributions to the conductivity.
The Luttinger interaction parameter for the new fermions (\ref{MT}) is $2K$.
As a result, the interaction among them vanishes at the Luther-Emery point,\cite{Luther74} $K=\frac{1}{2}$. 
Later, in Section~\ref{sec:InhomSystem} we  
will apply the above formalism to the 
wire with the leads, and develop a perturbative expansion around this special value for the Luttinger parameter within the leads, $K_L$.
 
In terms of the spinless fermionic degrees of freedom the charge current density, $J = e j$, is given by 
\begin{align} \label{J-e}
J = \sqrt{2} e K u \frac{\partial_x \theta_c}{\pi} 
= \sqrt{2}\,  e_{\rm CI}  \Psi^\dag v_J \sigma^3 \Psi.
\end{align} 
Again, the overall multiplicative factor $\sqrt{2}$ can be traced back to the spin degree of freedom of the original electrons. 
As announced, we absorbed the topological charge associated with the fermions, $Q=1/\sqrt{2}$, into the fractional electric charge quantum, $e_{\rm CI} = Q e$, of Eq.~(\ref{FracCharge}).
%
%
%
%
Moreover, the charge velocity of the new fermions reads
\begin{align} \label{Anomaly}
v_J = v \left(1 + \nu g_4 - \nu g_2 \right).
\end{align}
For the general case $g_2\neq g_4$, the velocity $v_J$ depends on the 
interactions (cf. Ref.~\onlinecite{GiamarchiBook}, Sec. 7.2).

The quadratic part of the refermionized theory (\ref{MT}) can be diagonalized by a Bogoliubov transformation resulting in the two bands with the spectrum $\pm \epsilon_k$, where $\epsilon_k = \sqrt{\Delta^2 +(v k)^2}$. When the chemical potential lies within the gap, $|\mu| < \Delta$, the ground state is a Mott insulator. 
When the chemical potential is tuned into one of the two bands the ground state becomes a conductor via the CI transition.

\subsection{Critical theory of the commensurate-incommensurate Mott transition}
\label{sec:CriticalTheory}
 
From now on in this work we will focus on the properties of the system close to the Mott transition and assume in the following that the chemical potential of the $\Psi$ fermions is near the upper band edge, 
$\mu \approx \Delta$, i.e., $r \approx 0$, where $r$ is the tuning parameter for the 
quantum phase transition, cf.~Eq.~(\ref{-r}).
The generalizaton to the situation when $\mu \approx -\Delta$ is straightforward and will not be further discussed.
 
The effective Hamiltonian governing the transition is given by nonrelativistic 
free spinless fermions,\cite{Japaridze78,Pokrovsky79}
\begin{align} \label{CI-theory}
\mathcal{H}_{\rm CI} =& 
\int dx\, c^\dag(x) \left[r - \frac{\partial_x^2}{2m}\right] c(x),
\end{align}
where the operator $c^\dag$ creates a fermionic excitation in the upper band, the mass is $m = \Delta/v^2$, and $r$ is defined in Eq.~(\ref{-r}). The scaling dimension of the tuning parameter 
and the critical dynamical exponent are 
dim$\,[r] = 1/\nu_r = 2$ and $z=2$, respectively. 

As the gas of fermions is very dilute at the transition, their interaction is weak. As the CI transition is approached, $r\to 0$, the interaction among the fermions decreases faster than their density, rendering the $c$ fermions free at the transition. Speaking crudely, the fermions are free not because they do not interact {\it per se} but because there are too few particles to interact with. Formally, this can be seen from considering the residual interaction
\begin{align} \label{CIInteraction}
\mathcal{H}^{\rm CI}_{\rm int} = \frac{1}{4} 
\sum_{k k' p p'} \delta_{k+k',p+p'} 
\Gamma^\pdag_{k k';p p'} c^\dag_{k} c^\dag_{k'} c^\pdag_{p} c^\pdag_{p'}. 
\end{align}
The interaction amplitude
$\Gamma_{kk';pp'}$ is strongly constrained as the Pauli principle requires the effective amplitude to be antisymmetric upon exchange of
momenta $p \leftrightarrow p'$ and $k \leftrightarrow k'$. Near the CI
transition we can expand the amplitude, and the lowest order contribution
is suppressed by two powers of momenta,
\begin{align} 
\Gamma^\pdag_{k k';p p'} = \gamma  (k-k') (p-p'),
\end{align}
with the coefficient $\gamma = - g_2 v^2/(4 \Delta^2)$. As a consequence, at the quantum critical point the residual interaction, $\gamma$, is irrelevant in the 
renormalization group sense. 
  
\begin{figure}
\includegraphics[width=3.4in]{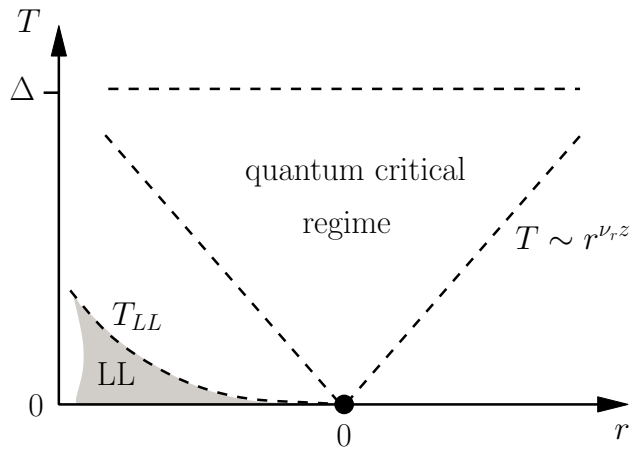}
\caption{\label{fig:1} Phase diagram of the CI transition for temperatures smaller than the gap $\Delta$, see text. A quantum phase transition with exponents 
$\nu_r = 1/2$ and $z = 2$ separates the commensurate, $r>0$, and the incommensurate, $r<0$, phases. The CI Hamiltonian (\ref{CI-theory}) controls the vicinity of the quantum critical point. The quantum critical regime occurs within a cone
$T\sim |r|^{\nu_r z} = |r|$.}
\end{figure}

The phase diagram of the CI Mott transition is shown in Fig.~\ref{fig:1}. The tuning parameter $r$ controls the distance to the quantum critical point; for $r>0$ the system is in the commensurate phase with an insulating ground state. 
In the metallic phase, $r<0$, at lowest temperatures $T<T_{\rm LL}$, with $\log(|r|/T_{\rm LL}) \propto 1/|r|$, a crossover between the critical behavior and the 
Luttinger liquid takes place; the residual interaction $\gamma$ at finite density
induces Luttinger liquid correlations in the spectral function of the $c$ fermions with the Luttinger parameter $K_c-1 \propto \sqrt{|r|}$. 

\subsection{Transport}
\label{sec:Transport}

We will consider electric transport through the critical Mott wire. Like the thermodynamic properties, the transport has universal characteristics at the Mott transition, on which we focus in the following. 
The conductance can be obtained 
from the zero frequency limit of the conductivity,\cite{FisherLee81}
\begin{align} \label{G-sigma}
G = \lim_{\omega\to 0} \sigma(x,x';\omega),
\end{align}
at fixed positions $x$ and $x'$. The conductivity $\sigma(x,x';\omega)$ describes the linear response of the charge current at position $x$ upon applying an 
external\cite{footnote-cond} 
electric field $\mathcal{E}(x',\omega)$ at position $x'$ with frequency $\omega$
\begin{align}
\langle J(x,\omega)\rangle = \int dx' \sigma(x,x';\omega) \mathcal{E}(x',\omega).
\end{align} 
The conductivity is derived with the help of the Kubo formula that relates $\sigma$ to the charge current autocorrelation function 
\begin{align} \label{kubo}
\sigma(x,x';\omega) = \frac{\K^{\rm ret}_{x,x';\omega} - \K^{\rm ret}_{x,x';\omega=0}}{i \omega} ,
\end{align}
where the retarded function is defined by 
\begin{align}
\K^{\rm ret}_{x,x';\omega} = i \int dt\, e^{i \omega t} \langle \left[ J(x,t), J(x',0)\right]\rangle .
\end{align}
The subtraction of the $\omega=0$ mode in (\ref{kubo}) ensures gauge invariance.
In the following, it will be convenient to obtain the retarded response function from the temperature correlation function
\begin{align} \label{CurrentAutoCorr}
\mathcal{K}_{x,x';\Omega_n} = \int_0^{1/T} d\tau\, e^{i \Omega_n \tau} \langle T_\tau J(x,\tau) J(x',0)\rangle 
\end{align}
by the analytic continuation from the upper half-plane, $\K^{\rm ret}_{x,x';\omega} = \mathcal{K}_{x,x';\Omega_n}|_{i \Omega_n \to \omega + i 0}$, where $\Omega_n = 2 \pi n T$ is the bosonic Matsubara frequency.

\subsection{Critical conductance of the homogeneous system}
\label{sec:HomSystem}

Consider first the uniform system close to the metal-insulator transition. 
At the criticality, the transport is determined by the free nonrelativistic 
$c$-fermions, i.e., by the Hamiltonian (\ref{CI-theory}). 
The resulting conductance 
\begin{align} \label{HomConductance}
G^{(\rm hom)} = \frac{2 e^2_{\rm CI}}{h} f(r/T) = \frac12 {G_0}  f(r/T)
\end{align}
is simply the conductance quantum $2e_{\rm CI}^2/h$
for the effective fermions with charge (\ref{FracCharge}), times the 
occupation factor that is given by the Fermi distribution (\ref{FermiFunction}). 

The critical conductance is universal and its temperature dependence is controlled by a Fermi function (\ref{FermiFunction}).
The non-universal corrections arising from the residual interaction (\ref{CIInteraction}) vanish with the distance to the quantum critical point as the interaction $\gamma$ is an irrelevant perturbation. In particular, in the limit of low temperatures the conductance at criticality, $r=0$, approaches the value
\begin{align}
\left.G^{(\rm hom)}\right|_{r=0, T\to 0} = \frac{1}{4} G_0 \,.
\end{align}
At zero temperature the conductance 
$G^{(\rm hom)}|_{T=0, r\to 0_+} \equiv 0$ in the insulating phase, 
while in the metallic phase $G^{(\rm hom)}|_{T=0, r\to 0_-} = \frac12 G_0$,
thus the conductance exhibits a jump across the transition of the height
\begin{align} \label{HomCondJump}
\left.G^{(\rm hom)}\right|_{T=0,r\to 0-} - \left.G^{(\rm hom)}\right|_{T=0,r\to 0+} = \frac{1}{2} G_0 \,.
\end{align}
The factor $\frac{1}{2}$ originates from the fractional nature of the charge carried by the critical fermionic degrees of freedom, $e_{\rm CI} = e/\sqrt{2}$. As was pointed out in Ref.~\onlinecite{Kolomeisky93}, the conductance jump across the transition (\ref{HomCondJump}) in fact corresponds to a full conductance quantum but with fractional charge, $G_0/2 = 2 e^2_{\rm CI}/h$.

In the following section the modifications to the critical conductance (\ref{HomConductance}) due to the presence of leads attached to the Mott wire are discussed.

\section{Critical conductance in the presence of leads}
\label{sec:InhomSystem}

In a two-terminal conductance measurement the critical Mott wire will
be connected to leads. In the presence of leads the conductance will
differ from the expression (\ref{HomConductance}) for the homogeneous
system. We assume here that the leads contain Luttinger liquids
characterized by the Luttinger parameter $K_L$. Here we argue that the
critical conductance is that of the homogeneous case,
Eq.~(\ref{HomConductance}), {\it only} if the Luttinger parameter
$K_L$ has the Luther-Emery value $K_L=1/2$.  Note that this value of
$K_L$ corresponds to {\it strong} interaction between the original
electrons in the leads. If $K_L$ differs from $\half$, the universal
properties of the critical conductance are modified. Below we develop
the perturbation theory in the deviation from the Luther-Emery point,
and evaluate the critical conductance in the first order in
$K_L-\frac{1}{2}$.

\subsection{Leads with Luttinger parameter $K_L=\frac{1}{2}$}
\label{sec:Leads1/2}

The $\Psi$ fermions entering the Thirring model (\ref{MT}) are the appropriate degrees of freedom describing the Mott transition as the interaction, $g_2$ and $g_4$, among them is irrelevant. In the following we will also adopt  a description of the leads, where the Mott gap is absent, $\Delta=0$, in terms of the same $\Psi$-fermions. The advantage of such a description is that the Mott region of the wire can be incorporated into an effective scattering approach for the $\Psi$ fermions which is outlined in the following. 
The caveat is that, unfortunately, these effective fermions will in general, i.e.~for $K_L \neq \half$, interact within the leads. This is what makes this 
problem difficult, as one has to solve the strongly interacting fermionic problem
which also lacks translation invariance.
The influence of the  interaction within the leads
is considered in Section \ref{subsec:LutherEmeryCorr} perturbatively.

Mori {\it et~al.} in Ref.~\onlinecite{Mori97} considered a very similar setup as we do here. They also calculated the current-current correlation function with the help of the $\Psi$ fermions at the Luther-Emery point but only for the uniform system, 
where the Mott gap extends over the full wire. 
 They proceeded by using the expression for the homogeneous case to derive the conductance in the presence of the leads where the original electrons do not interact. We think that the last step has led them to the erroneous result for the temperature dependence of the critical conductance. We will further elaborate on the difference between the treatment of Mori {\it et al.} and ours in Section \ref{sec:Summary}.

\subsubsection{Scattering states}

Let us  consider the situation where the Luttinger parameter of the leads 
$K_L = \frac{1}{2}$, corresponding to $g_2\equiv 0$ in the Hamiltonian (\ref{MT}). 
Here we also set $g_4=0$.
The absence of interactions allows us to describe the wire in terms of a simple 1D scattering problem. The leads can then be characterized by the states of the $\Psi$ fermions that scatter off the Mott region. The crucial point is that the latter will be treated as an effective point scatterer. This is a correct approximation
for the dc Landauer transport, and physically it means that we attach infinitely
long leads to the Mott region.

The scattering wave incident from the left has the standard form
\begin{subequations} \label{wf}
\begin{align} \label{ScatteringStatesL}
\lefteqn{\psi_{\epsilon,+}(x) =}
\\\nn&
\frac{1}{\sqrt{2\pi v}}
\left\{
\begin{array}{ll}
e^{i \epsilon x/v} \begin{pmatrix} 1 \\ 0 \end{pmatrix} + 
e^{- i \epsilon x/v} r_{L}(\epsilon) \begin{pmatrix} 0 \\ 1 \end{pmatrix},
& x < 0 
\\
e^{i \epsilon x/v} t_{L}(\epsilon) \begin{pmatrix} 1 \\ 0 \end{pmatrix},
& x > 0.
\end{array}
\right.
\end{align}
The energy $\epsilon$ is measured here (and everywhere below) with respect to the Fermi level. The phase velocity in the 
R- and L- moving states (upper and lower spinor components)
equals $\pm v$ correspondingly.
The wave incident from the right reads
\begin{align}\label{ScatteringStatesR}
 \lefteqn{\psi_{\epsilon,-}(x) =}
\\\nn&
\frac{1}{\sqrt{2\pi v}}
 \left\{
\begin{array}{ll}
e^{-i \epsilon x/v} t_{R}(\epsilon) \begin{pmatrix} 0 \\ 1 \end{pmatrix},
& x < 0 
\\
e^{-i \epsilon x/v} \begin{pmatrix} 0 \\ 1 \end{pmatrix} + 
e^{i \epsilon x/v} r_{R}(\epsilon) \begin{pmatrix} 1 \\ 0 \end{pmatrix},
& x > 0.
\end{array}
\right.
\end{align}
\end{subequations}
The transmission and reflection coefficients are components of the  
scattering S-matrix,
\begin{align} \label{SMatrix}
\mathbf{S}(\epsilon) = 
\begin{pmatrix}
t_L(\epsilon) & r_R(\epsilon) \\
r_L(\epsilon) & t_R(\epsilon)
\end{pmatrix},
\end{align}
which is unitary, $\mathbf{S}^\dag \mathbf{S} = \mathds{1}$. 
The unitarity of the S-matrix ensures that it has only three independent parameters,
which can be chosen in the form
\begin{align}
t_L = t_R = {\sf t}\, e^{i\phi_t},\quad
r_L = - {\sf r}\, e^{i\phi_-},\quad
r_R = {\sf r}\, e^{i\phi_+}
\end{align}
with $\phi_t = (\phi_+ + \phi_-)/2$, and positive ${\sf r}$ and ${\sf t}$, ${\sf r}^2+{\sf t}^2=1$. The relation $t_L=t_R$ is a consequence of the time-reversal symmetry.

The presence of the Mott gap at the center of the wire makes the transmission and the reflection coefficients strongly energy-dependent: The propagation is blocked for the fermionic states with energies within the gap, $\epsilon < r$. These states therefore have zero transmission coefficient as we assume the 
Mott gapped region of the wire to be sufficiently long such that the tunneling across it can be neglected. On the other hand, the fermionic states with energies above the gap, $\epsilon > r$, are able to traverse the Mott region. We will assume that the leads are attached to the Mott-gapped region of the wire in the adiabatic fashion 
(i.e. $V_{\rm Umkl}=V_{\rm Umkl}(x)$ is a smooth function), 
such that the backscattering of particles moving above the gap can be neglected. The tuning parameter, $r$, of the transition thus divides the fermionic scattering states into the propagating waves with unit transmission for energies $\epsilon > r$, and into the standing waves for energies $\epsilon < r$. As a result, we model the Mott region of the wire by utilizing simple unit step-function transmission and reflection amplitudes
\be
{\sf t(\epsilon)} = \Theta(\epsilon-r) \,, \quad 
{\sf r(\epsilon)}=\Theta(r-\epsilon) \,.
\ee
Summarizing,
\begin{align} \label{SMatrix-Theta}
\mathbf{S}(\epsilon) = 
\begin{pmatrix}
\Theta(\epsilon-r) e^{i\phi_t(\epsilon)} & \Theta(r-\epsilon) e^{i\phi_+(\epsilon)}
\\
-\Theta(r-\epsilon) e^{i\phi_-(\epsilon)} & \Theta(\epsilon-r) e^{i\phi_t(\epsilon)}
\end{pmatrix} .
\end{align}
%
%
%
The scattering phases $\phi_{\pm}(\epsilon)$  
are accumulated while traversing or reflecting from the Mott barrier. 
The barrier shape is encoded in their smooth energy dependencies,
\begin{align} \label{Phases}
\phi_{\pm}(\epsilon) \simeq \phi_{\pm}^0 + \ell_{\pm} \epsilon/v, \quad
\end{align}
The scattering lengths $\ell_{\pm}$ are a measure of the effective 
length of the region of the wire where the Mott gap is present.

\subsubsection{Green's Function}

%
%
%

The single-particle {\it Green's function} in the Matsubara representation 
is a $2\times2$ matrix in the spinor space [$a,b = 1,2$ are the spinor components
of the wave functions (\ref{wf})]
\bse \label{Gxx'}
\be \label{Gxx't} 
\G^{ab}_{x,x';\tau} = - \la T_\tau \psi^a(x,\tau) \psi^{\dag b}(x',0)\ra .
\ee
In what follows we will be using its Lehmann representation
\bea  
\label{G-lehmann}
\G_{x,x';\omega_n} &=&
\int\! d{\epsilon}\, {\mathbf{A}_{x,x';\epsilon} \over i\omega_n-\epsilon} \,, \\
\mbox{where }\quad \mathbf{A}^{ab}_{x,x';\epsilon} &=& \sum_{\sigma=\pm}
\psi^a_{\epsilon,\sigma}(x) \psi^{\dag b}_{\epsilon,\sigma}(x')\,,
\label{Aab}
\eea
\ese
and $\omega_n = \pi (2n+1) T$ is a fermionic Matsubara frequency.

As we show now, 
the Green's function (\ref{Gxx'}) has qualitatively different properties
when its arguments $x$ and $x'$ are located on the same and on the 
opposite sides of the impermeable potential barrier. 
Introducing the labels
\be \label{ss'}
s=\sign x \quad \mbox{and}\quad s'=\sign x' \,,
\ee
one finds:
\bse \label{A}
\begin{align}
\lefteqn{s\neq s':}\nn \\
& 
\mathbf{A}_{x,x';\epsilon} = \nu \Theta_{\epsilon-r}
\begin{pmatrix} e^{i\epsilon(x-x')/v+ i s \phi_t(\epsilon)} & 0 \\ 0 & e^{-i\epsilon(x-x')/v - i s \phi_t(\epsilon)}\end{pmatrix}
\label{A-opposite}
\\ 
\lefteqn{s=s':} \nn
\\ &
\mathbf{A}_{x,x';\epsilon} =
\nu 
\begin{pmatrix} e^{i\epsilon(x-x')/v} & 0 \\ 0 & e^{-i\epsilon(x-x')/v}\end{pmatrix}
\nn
\\
&\quad+ \nu s\, \Theta_{r-\epsilon}
\begin{pmatrix} 0 & e^{i\epsilon(x+x')/v + i s \phi_s(\epsilon)} \\ 
e^{-i\epsilon(x+x')/v - i s \phi_s(\epsilon)} & 0 \end{pmatrix}\!. \quad\quad
\label{A-same}
\end{align}
\ese
The spectral function (\ref{A-opposite}), as well as the 
first term in the spectral function (\ref{A-same}) 
become translation-invariant in the limit $|x-x'|\to \infty$, since they both
depend only on the coordinate difference $x-x'$. 
They originate from the propagating states 
and from the corresponding co-moving parts of the standing waves.
The second term in Eq.~(\ref{A-same}) is due to the counter-moving parts of the standing waves and  
comes only from the states
with energies below the barrier, $\epsilon<r$. It lacks translational invariance,
and is affected by the barrier shape through the energy-dependent
reflection phase shifts $\phi_\pm$.


\subsubsection{Polarization operator}

A basic ingredient in the following calculation is the polarization operator 
that is a convolution of the temperature Green functions (\ref{Gxx'}),
\be \label{Pi-ij-def}
\Pi_{x,x';\W_n}^{ij} = 
-T\sum_{\omega_n} \tr \lf \tau^i \G(x,x';\w_n+\W_n)\tau^j \G(x',x;\w_n) \rf
\ee
where $i,j=0,1$ and 
\be \label{SigmaMatrices}
\tau^i = \begin{pmatrix} 1 & 0 \\ 0 & \beta^i \end{pmatrix} ,
\quad \mbox{with} \quad
\beta^i = \lf \begin{matrix} +1\,, \quad i = 0\,; \\ 
-1\,, \quad i=1\,. \end{matrix} \right.
\ee
The polarization matrix $\Pi_{x,x';\W_n}^{ij}$ 
depends on the bosonic Matsubara frequency $\W_n = 2 \pi n T$. A detailed calculation of the above expression can be found in the Appendix \ref{app:pol}. 
In the dc limit ($i\W_n \to \omega + i0$, $\omega\to 0$) it reduces to
%
\begin{align} \label{Pi-Pi0}
\Pi^{ij}_{x,x';\W_n} &= \Pi^{(0)\,ij}_{x,x';\W_n} + \lb f(r/T)-1 \rb
\\\nn
&\times \lf {1-ss'\over 2} \Pi^{(0)\,ij}_{x,x';\W_n} 
-\beta^j {1+ss'\over 2} \Pi^{(0)\,ij}_{x,-x';\Omega_n}
\rf,
\end{align}
where $f$ is the Fermi function (\ref{FermiFunction}), and we used the 
abbreviation (\ref{ss'}).
The polarization operator matrix in the absence of the Mott barrier is
\bea \label{Pi0-1}
\Pi^{(0)\,ij}_{x,x';\W_n} &=& 2 \nu \delta_{ij} \delta(x-x') + \bar{\Pi}^{ij}_{x,x';\W_n} \,, \\
\label{Pi0-2}
\bar{\Pi}_{x,x';\W_n} &=& -2 \pi \nu^2 e^{-|\Omega_n||x-x'|/v} \\\nn
&\times& 
\begin{pmatrix} |\W_n| & \W_n \sign(x-x') \\ \W_n \sign(x-x') & |\W_n|
\end{pmatrix},
\eea
with the density of states (\ref{nu}). The rows and columns of the 
matrix (\ref{Pi0-2}) are numbered by the indices $i,j=0,1$, such that 
$\bar\Pi^{00}$ corresponds to the upper left corner.
The corrections to the result (\ref{Pi-Pi0}) are subleading in powers of frequency $\Omega_n$, see Appendix \ref{app:pol} for details. 
 
Let us discuss the meaning of the expression (\ref{Pi-Pi0}).
The trivial observation 
is that, in the absence of the barrier, $f= 1$, and $\Pi\equiv \Pi^{(0)}$
becomes translation invariant.
In the opposite case of an infinitely high barrier, $f=0$,
and $\Pi^{ij}_{x,x';\W_n} \equiv 0$ identically vanishes when the 
points $x$ and $x'$ are on the opposite sides of the barrier (no transmission).
We now consider the polarization operator when $x$ and $x'$ are on the same side, 
$\Pi^{ij}_{x,x';\W_n}=\Pi^{(0) ij}_{x,x';\W_n} + \beta^j \Pi^{(0) ij}_{x,-x';\W_n}$.
This expression has a static part (that remains finite in the limit $\W_n \to 0$), 
and the dynamic part $\propto \W_n$.
The static part of the compressibility 
$\Pi^{00}_{x,x';\W_n=0} = 2\nu \delta(x-x')$ is not affected by the barrier
at all (this in fact is true for any value of $f$), 
as the density of states in the leads is independent of the barrier. 
The dynamic part is more interesting. For $f=0$, it is given by
$\bar\Pi^{ij}_{x,x';\W_n} + \beta^j \bar\Pi^{ij}_{x,-x';\W_n}$.
This expression can be understood by looking at the dynamic part of the 
current-current correlator, corresponding to $i=j=1$.
After the analytic continuation it behaves  $\propto \w e^{i\w x/v}\sin(\w x'/v)$.
Therefore, the conductivity (\ref{kubo}) {\it vanishes} as $\sin(\w x'/v)$
in the dc limit: Given enough time, the system finds out about the barrier.
For finite $f$, the part of the expression (\ref{Pi-Pi0}) responsible for 
the transmission of the particles is proportional to $f(r/T)$.

Technically, the presence of the barrier leads to an additional term in the expression (\ref{Pi-Pi0}) that is multiplied by the factor $f(r/T) - 1$. This term can be further simplified by noting that, in the dc limit, the point-scatterer approximation for the 
Mott region allows us to substitute
$|x+x'|=|x|+|x'|$ for $s=s'$, and $|x-x'|=|x|+|x'|$ for $s\neq s'$, yielding
\bea
\Pi^{(0) ij}_{x,x';\W_n} &=& \bar \Pi^{ij}_{x,0;\W_n} e^{-|\W_n x'|/v} \,, \quad s\neq s', \quad\\
\Pi^{(0) ij}_{x,-x';\W_n} &=& \bar \Pi^{ij}_{x,0;\W_n} e^{-|\W_n x'|/v} \,, \quad s=s' .
\eea
Using the above simplification the expression Eq.~(\ref{Pi-Pi0}) 
can be cast into a form characteristic for a scattering problem, 
\be \label{Polarizations}
\Pi_{x,x';\W_n} = \Pi^{(0)}_{x,x';\W_n} + \bar\Pi_{x,0;\W_n}\,\what{\mathfrak{T}}\,\bar\Pi_{0,x';\W_n}.
\ee
In the limit of small $\W_n$ 
(in the sense of analytic continuation $i\W_n \to \w\to 0$ specified above),
the scattering $\what{\mathfrak{T}}$-matrix is given by
\be \label{BosonicTMatrix}
\what{\mathfrak{T}} = \begin{pmatrix} 0 & 0 \\ 0 & \mathfrak{T} \end{pmatrix} ,
\quad \mathfrak{T} = \frac{1-f(r/T)}{2 \pi \nu^2 |\Omega_n|} \,.
\ee

\subsubsection{Critical conductance with leads at the Luther-Emery point $K_L = \frac{1}{2}$}
\label{subsubsec:LutherEmery}

From the evaluated polarization matrix (\ref{Polarizations}) we can derive the conductance in the presence of leads that have a Luttinger parameter of the Luther-Emery value $K_L=\frac{1}{2}$. The current autocorrelator (\ref{CurrentAutoCorr}) is then given by 
\begin{align}
\mathcal{K}^{\rm LE}_{x,x';\Omega_n} = 2 e^2_{\rm CI} v^2 \Pi^{11}_{x,x';\Omega_n}.
\end{align}
The conductance (\ref{G-sigma})
follows from the Kubo formula (\ref{kubo}), 
\begin{align} \label{CriticalLLConductance}
G_{\rm LE} = \frac{1}{2} G_0\,  f(r/T),
\end{align}
with the conductance quantum $G_0 = 2 e^2/h$. We thus reproduce the critical conductance of the homogeneous system (\ref{HomConductance}). Below, we show how the critical properties of the conductance are modified in first order in the deviation $K_L-\frac{1}{2}$.

\subsection{Critical conductance: Perturbation theory around the Luther-Emery point}
\label{subsec:LutherEmeryCorr}

\begin{figure}
(a)
\includegraphics[width= 0.22\linewidth]{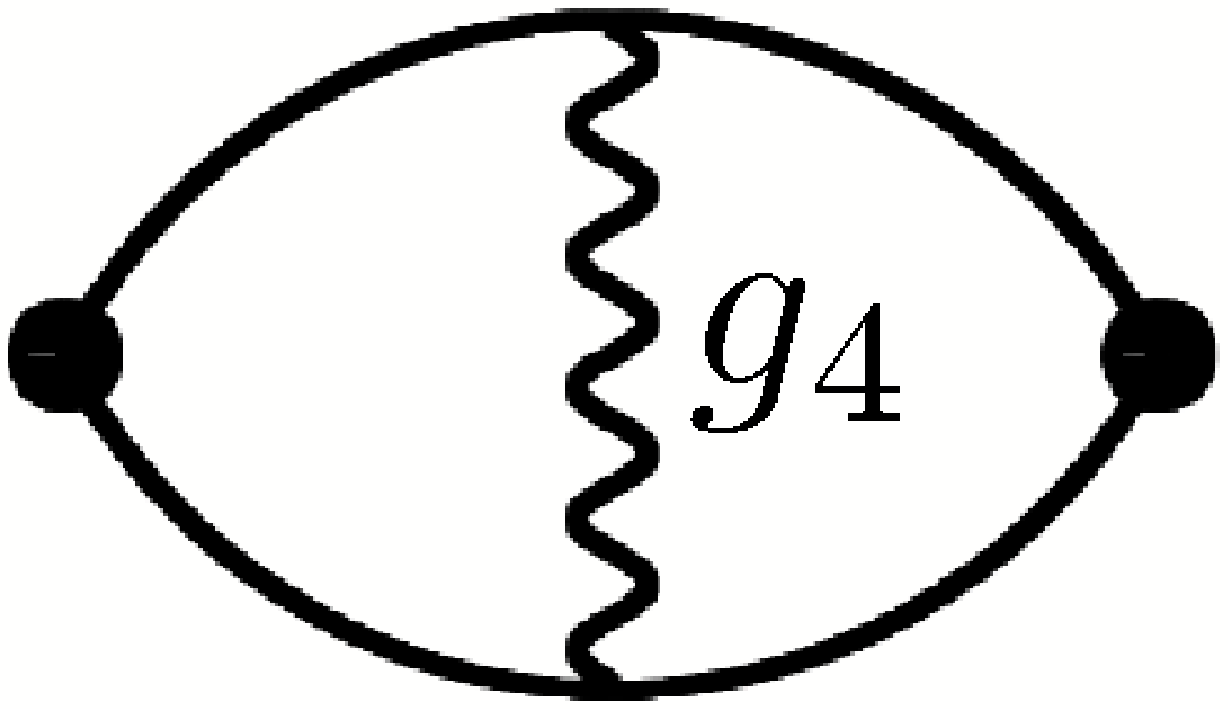}
\quad
\includegraphics[width= 0.5\linewidth]{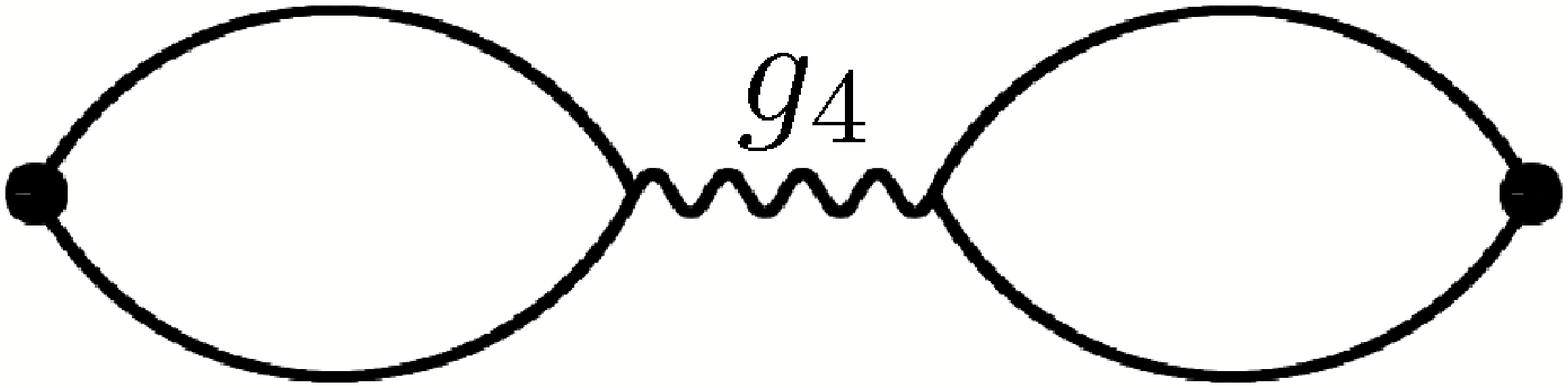}
(b)
\\[0.5em]
(c)
\includegraphics[width= 0.22\linewidth]{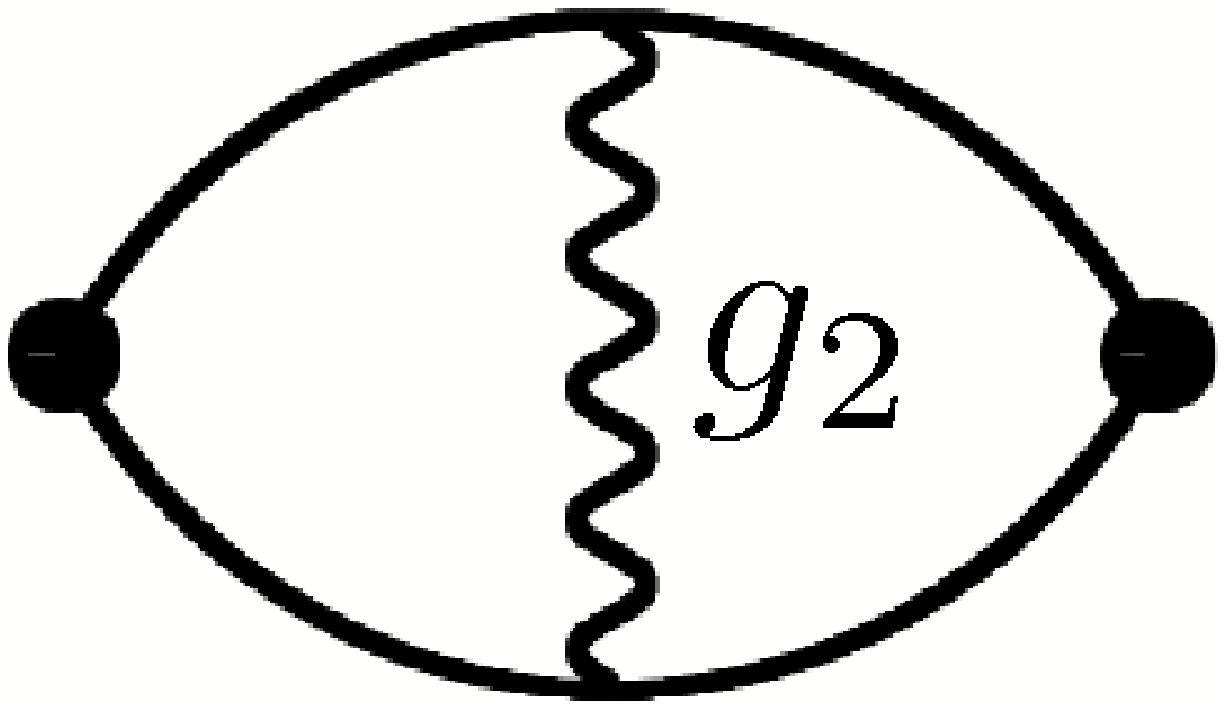}
\quad\quad
\includegraphics[width= 0.22\linewidth]{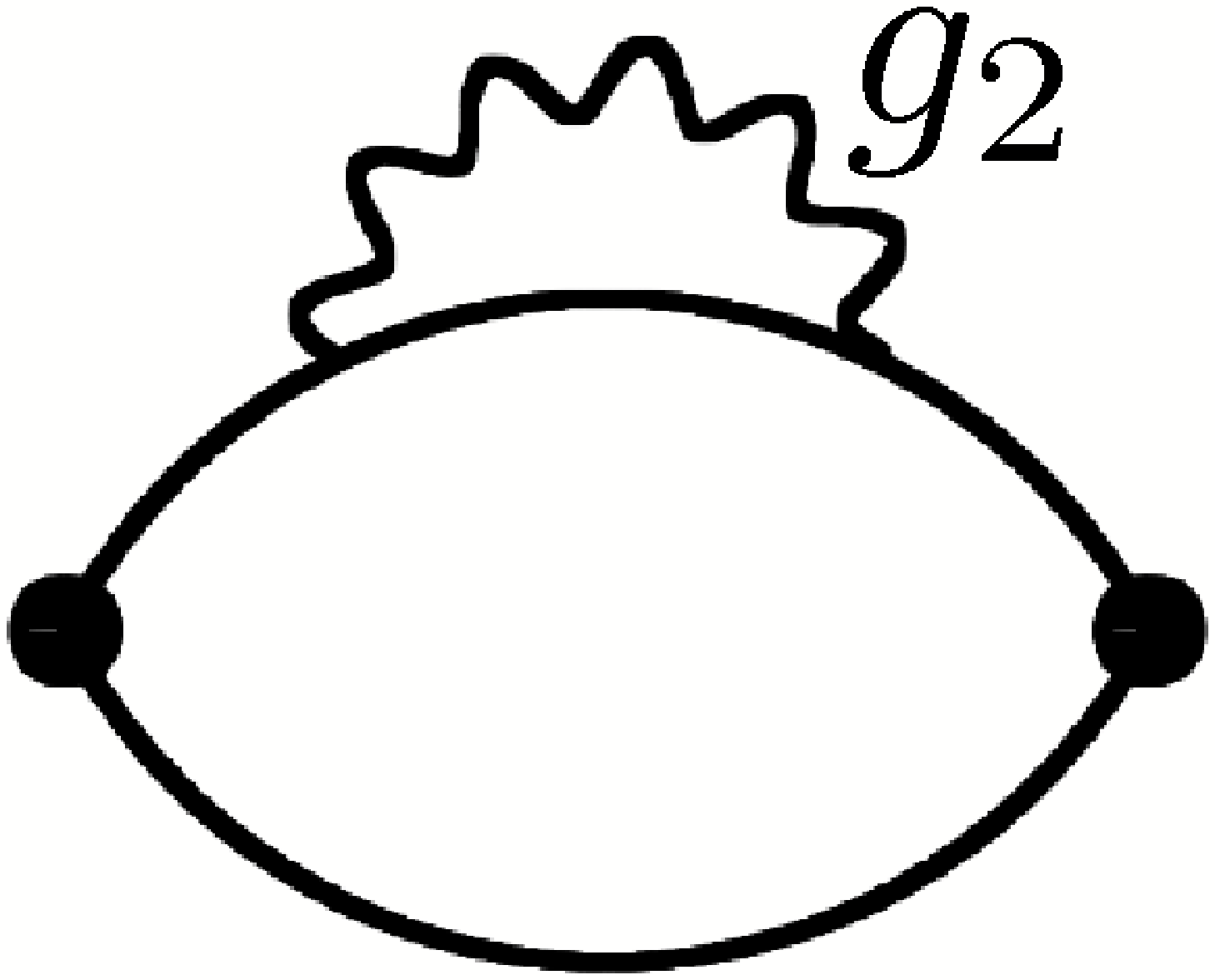}
(d)
\\[0.5em]
(e)
\includegraphics[width= 0.22\linewidth]{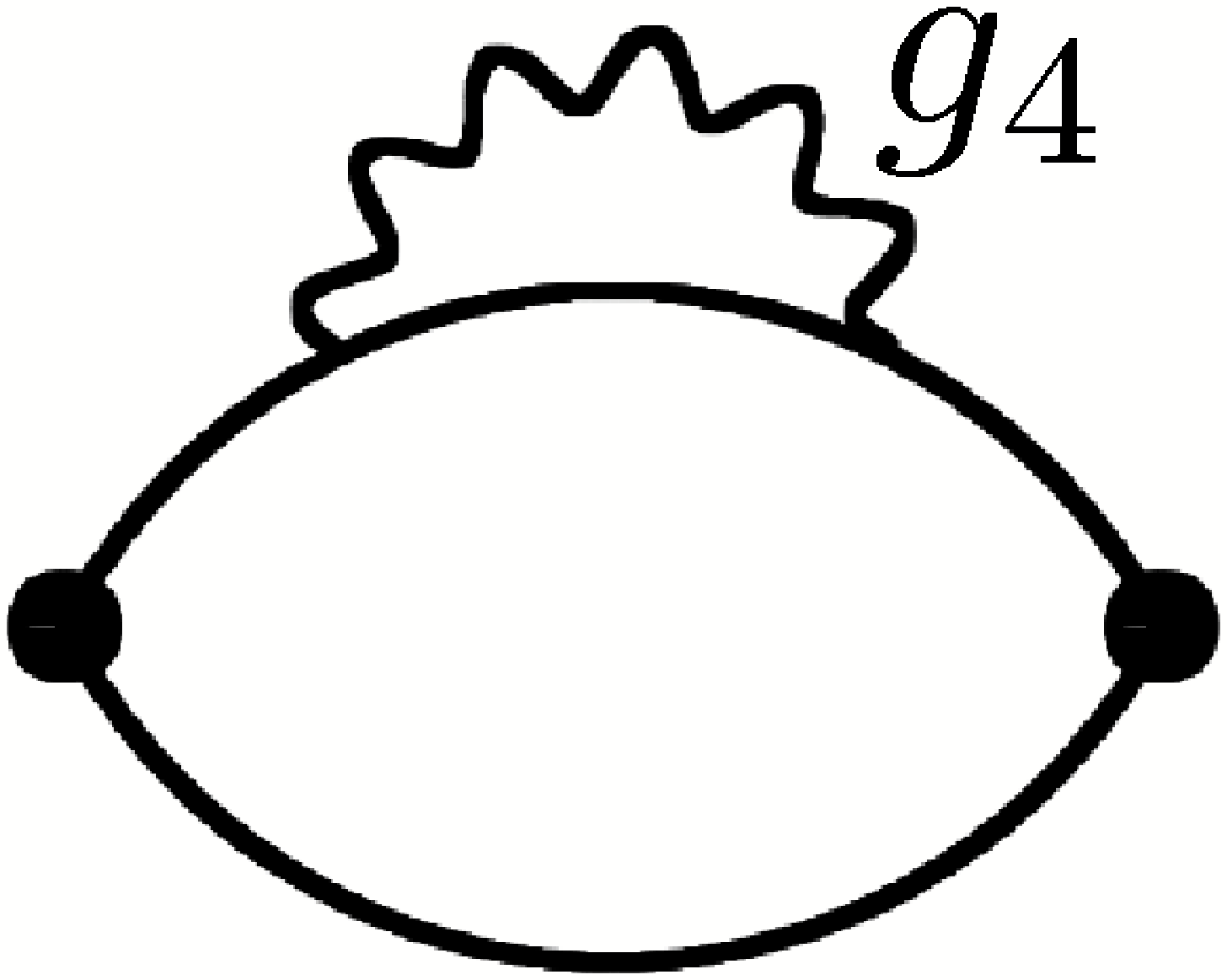}
\quad
\includegraphics[width= 0.5\linewidth]{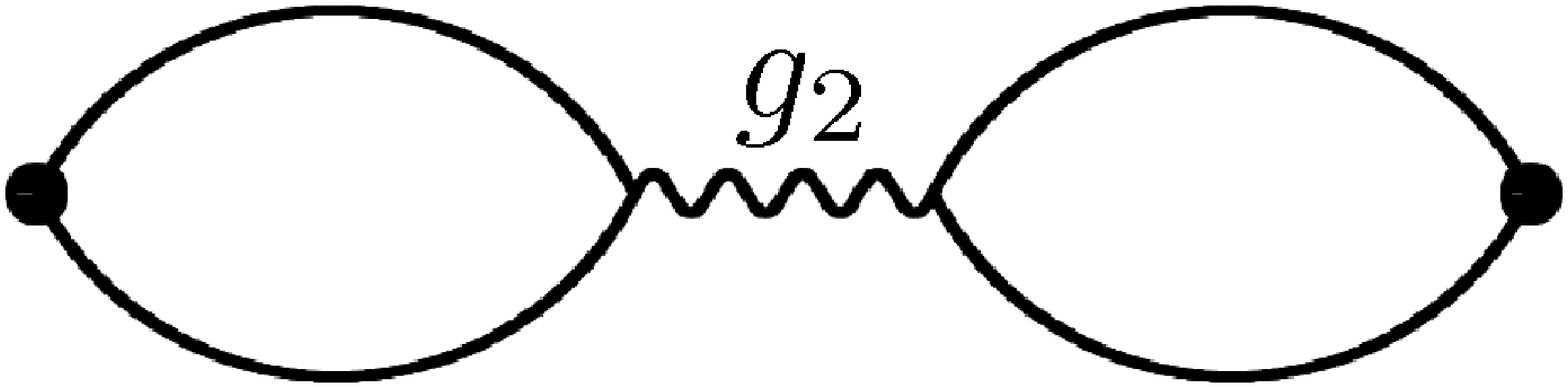}
(f)
\caption{\label{fig:Diagrams}  First order corrections to the current autocorrelator. The straight line represent fermionic $\Psi$ propagators, the wiggled line is the interaction and the dot is the current vertex $\sqrt{2} e _{\rm CI} v \sigma^3$.  }
\end{figure}

Here we evaluate the correction to the conductance arising from the deviation of the Luttinger parameter in the leads from its Luther-Emery value $K_L=\frac{1}{2}$. The deviation from $K_L=\frac{1}{2}$ implies according to Eq.~(\ref{EffLuttingerParameters2}) a finite $g_2$ interaction between the scattering $\Psi$ fermions. In the following we calculate the conductance up to the first order in the interactions $g_2$ and $g_4$. As in the first order the $g_4$ interaction only modifies the velocity, it is expected that it does not enter the conductance correction. The verification of this expectation serves as a check of our calculation and is actually the only reason to introduce at all the redundant $g_4$ interaction in the mapping (\ref{EffLuttingerParameters1}) and (\ref{EffLuttingerParameters2}). 

There are various contributions to conductance arising from finite $g_2$ and $g_4$. First of all, these interactions modify the expression of the charge current velocity $v_J$, see Eq.~(\ref{Anomaly}). As a consequence, the conductance obtains the modification,
\begin{align}\label{correction-A}
\delta G^{\rm A} / G_{\rm LE} 
= 2 (g_4 \nu - g_2 \nu).
\end{align}
In addition, there are in total six first order diagrams that contribute to the current autocorrelator, see Fig.~\ref{fig:Diagrams}.
Diagrams (a) and (b) cancel each other exactly due to the Fermi statistics. 
Diagram (c) does not modify the conductance because its contribution to $\sigma(x,x';\omega)$ vanishes in the low frequency limit. 
In the absence of the barrier
this is self-evident, as the small frequencies correspond to small incoming 
momenta that cannot change an R-mover into a L-mover. In the presence of the barrier
this argument essentially goes through, since only the propagating 
parts (\ref{A-opposite}) of the Green's functions contribute; 
the latter have the same diagonal form as in the translation-invariant problem. 
Diagram (d) corresponds to the conductance correction coming from fermion scattering off a Friedel oscillation;\cite{Matveev93,Yue94,Nazarov03} we will see that in our case this contribution is absent.
The only modifications to the conductance arise from the self-energy diagram (e) and from the random-phase approximation (RPA) diagram (f). 
They are discussed in detail below.

\subsubsection{Scattering off Friedel oscillations: Diagram (d)} 


The self-energy correction in diagram (d) embodies the scattering off a Friedel oscillation. In the general case of a semitransparent barrier, this process leads to a logarithmic renormalization of the scattering coefficients and of the conductance.\cite{Matveev93,Yue94,Nazarov03} However, the S-matrix in our case has a very specific energy dependence (\ref{SMatrix-Theta}) as the transmission at a given energy is either zero or unity. In such a situation the renormalization e.g. of the transmission coefficient vanishes as it is proportional to the product of transmission and reflection coefficients at a given energy. As a consequence, diagram (d) does not contribute to the conductance here. 

The fact that there is no contribution from the Friedel oscillation 
in the first order in $g_2$ can be undestood already 
from the following simple argument.
Consider the propagating state, for which the transmission is unity in the absence
of interactions between fermions. Due to the interaction $g_2$ 
with the counterpropagating states, the 
reflection amplitude will appear, being proportional to $g_2$, thus the reflection
probability is ${\cal O}(g_2^2)$. This means that the 
transmission amplitude (and the probability) is of the form $1-{\cal O}(g_2^2)$, 
i.e. the correction to the transmission (and hence, to the conductance)
vanishes to the first order in $g_2$.

\subsubsection{Density of states renormalization: Diagram (e)}

The other fermionic self-energy that appears in diagram (e) in Fig.~\ref{fig:Diagrams} arises from the $g_4$ interaction. The left and right moving fermions interact only among themselves via $g_4$ which ensures that the resulting self-energy is translationally invariant,
\begin{align}
\what{\Sigma}^{(e)}(x,x') 
&= - \frac{g_4(x,x')}{2 \pi i(x-x')} \sigma^3,
\end{align}
as $g_4(x,x') = g_4(x-x')$. This self-energy causes a perturbative correction to the velocity of the scattering particles. This is seen from the analysis of the Schr\"odinger equation for the scattering states,
\begin{align}
- i v \sigma^3 \partial_x \psi_\epsilon(x)
+ \int dx' \what{\Sigma}^{(e)}(x,x') \psi_\epsilon(x') = \epsilon \psi_\epsilon(x).
\end{align}
The Schr\"odinger equation can be solved perturbatively with the help of the Wentzel-Kramers-Brillouin ansatz
\begin{align}
\psi_\epsilon(x) \propto \exp\left(i \sigma^3 \int^x dy\, \varphi(y)\right)
\end{align}
where in zeroth order $\varphi^{(0)} = \epsilon/v$. The first order correction to the eikonal phase is determined by
\begin{align}
\varphi^{(1)}(x) 
= \int dx' \frac{g_4(x,x')}{2 \pi v i(x-x')} e^{i \epsilon (x'-x)/v}.
\end{align}
In particular, the energy derivative of the eikonal phase up to first order is given by 
\begin{align}
\frac{\partial \varphi}{\partial \epsilon}
= \frac{1-\nu g_4}{v} 
\end{align}
where $g_4 = \int dx' g_4(x-x') e^{i \epsilon (x'-x)/v}$ is the Fourier transform of the $g_4$ interaction at small wavevector $\epsilon/v$.
This is equivalent to the renormalization of the velocity 
$v\to v+\delta v$, where $\delta v/v = g_4 \nu$.

The velocity renormalization enters the normalization of the scattering states (\ref{ScatteringStatesL}) and (\ref{ScatteringStatesR}) and thus contributes to the conductivity by changing the density of states,
%
$\nu \to \nu (1-\nu g_4)$.
%
The density of states on the other hand enters the polarization matrix (\ref{Polarizations}) which finally leads to the change in the conductance
\begin{align}\label{correction-e}
\delta G^{(e)}/G_{\rm LE} = -2 g_4 \nu
\end{align}
attributed to diagram (e).

\subsubsection{Random-phase approximation contribution: Diagram (f)}

The remaining RPA diagram (f) in Fig.~\ref{fig:Diagrams} can be reduced to the spatial convolutions of the two components of the polarization matrix (\ref{Pi-ij-def}). The resulting correction to the current autocorrelator (\ref{CurrentAutoCorr}) reads
\begin{align} \label{RPA-Current}
\lefteqn{
\delta \mathcal{K}^{(f)} (x,x';\Omega_n) =
}
\\\nn&
- g_2 e^2_{\rm CI} v^2\int d x_1
\left[
\Pi^{10}_{x,x_1;\Omega_n} \Pi^{01}_{x_1,x';\Omega_n} - 
\Pi^{11}_{x,x_1;\Omega_n} \Pi^{11}_{x_1,x';\Omega_n} \right]
\end{align}
As we are only interested in the behavior at small frequency $\Omega_n$, we can use the simplified expressions (\ref{Polarizations}) for the polarization operators. The integral over $x_1$ can then be performed with the help of the convolution (no summation over index $i$ implied)
\begin{align} \nn
\lefteqn{\int d x_1 \bar{\Pi}^{1i}_{x,x_1;\Omega_n} \bar{\Pi}^{i1}_{x_1,x';\Omega_n}}
\\ &
= \nu \bar{\Pi}^{11}_{x,x';\Omega_n} 
\left( \beta_i -  \frac{|\Omega_n|}{v} |x-x'| \right)
\end{align}
where $\beta_i$ was defined in Eq.~(\ref{SigmaMatrices}) and $\bar{\Pi}^{ij}$ in Eq.~(\ref{Pi0-2}).
After the evaluation of the $x_1$ integral, we obtain for the RPA contribution
\begin{align}
\delta \mathcal{K}^{(f)} (x,x';\Omega_n) &=
- 2 g_2\nu  e^2_{\rm CI} v^2\left[ 
- 2 \nu \delta(x-x') 
\right.
\\\nn&\left.
 - \bar{\Pi}^{11}_{x,x';\W_n} + 
\bar{\Pi}^{11}_{x,0;\W_n} \mathfrak{T}\, \bar{\Pi}^{11}_{0,0;\Omega_n} 
\mathfrak{T}\, \bar{\Pi}^{11}_{0,x';\W_n}
\right]
\end{align}
 The presence of the product of bosonic T-matrices, $\mathfrak{T}$, see Eq.~(\ref{BosonicTMatrix}), leads to the correction to conductance that contains an additional Fermi function,
\begin{align} \label{correction-f}
\delta G^{(f)}/G_{\rm LE} =  \nu g_2 \left[2 - f(r/T)\right].
\end{align}

\subsubsection{Full perturbative correction}

Collecting all the perturbative contributions to conductance, (\ref{correction-A}), (\ref{correction-e}) and (\ref{correction-f}), we obtain the following 
for the first order correction in the parameter $K_L-\frac{1}{2}$,
\begin{align}
\delta G/G_0  = - \frac{\nu g_2}{2} f^2(r/T) 
= \left(K_L-\half \right) f^2(r/T),
\end{align}
where we used the identification $\nu g_2 = 1 - 2K_L$ valid to the first order in $g_2$, cf. Eq.~(\ref{EffLuttingerParameters2}). 
As expected, the first order correction to conductance is independent of $g_4$.
Note that the correction starts with the square of the Fermi function whose origin can be traced back to the RPA diagram (f) in Fig.~\ref{fig:Diagrams}.
Adding to this the zeroth order contribution (\ref{CriticalLLConductance}) we finally obtain  the result 
announced in Eq.~(\ref{FinalResult}).

\section{Summary and discussion}
\label{sec:Summary}

\begin{figure}
\includegraphics[width= \linewidth]{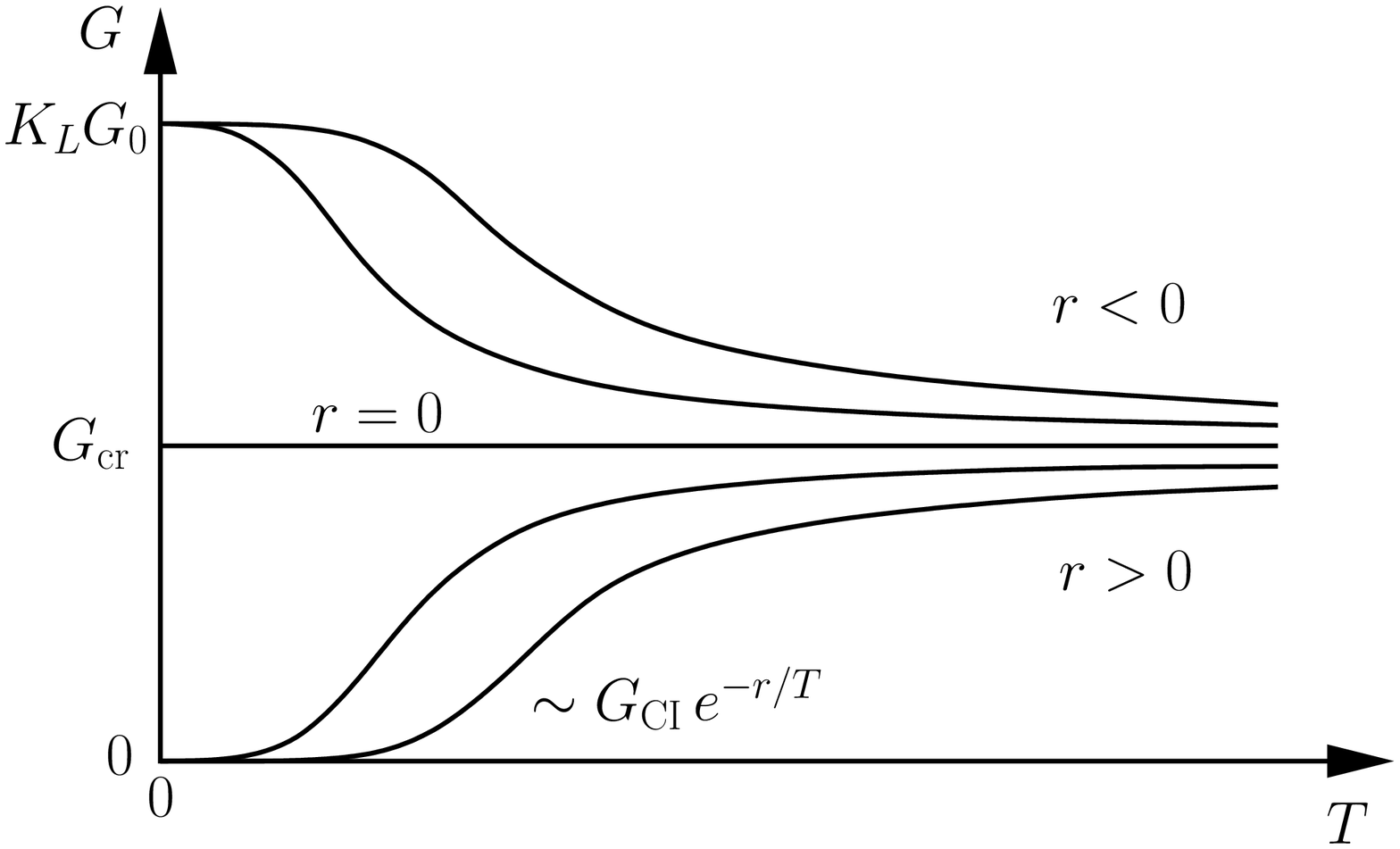}
\caption{\label{fig:2} Temperature dependence of two-terminal conductance close to the Mott transition for different values of the tuning parameter $r$. On the metallic side, $r<0$, at $T=0$ the conductance is determined by the Luttinger parameter of the leads, $K_L$; $G_0 = 2 e^2/h$ is the conductance quantum. In the insulating phase, $r>0$, at lowest temperature the conductance is thermally activated with a universal exponential prefactor involving the fractional charge, $e_{\rm CI} = e/\sqrt{2}$, of the critical degrees of freedom, $G_{\rm CI} = 2 e^2_{\rm CI}/h$. In the quantum critical regime, $|r| < T$, of the phase diagram, see Fig.~\ref{fig:1}, the conductance is determined by its critical value, $G_{\rm cr}$, see Eq.~(\ref{Gcr}), 
that depends both on the Luttinger parameter, $K_L$, and on 
the fractional charge, $e_{\rm CI}$.}
\end{figure}

In the present work we considered the finite-temperature dc conductance close to the commensurate-incommensurate Mott transition in a one-dimensional wire as a function of the tuning parameter, $r$, of the transition. For a homogeneous system in the absence of leads the result is universal and is given by Eq.~(\ref{HomConductance}). It is determined by the critical degrees of freedom of the transition --- the fermionic charge carriers that have a fractional charge $e_{\rm CI} = e/\sqrt{2}$, ---  
and the temperature dependence is simply a Fermi function. 

Furthermore, we addressed the question on how the conductance is modified when the leads are attached to the Mott wire. 
The result is summarized in Fig.~\ref{fig:2}. In the incommensurate phase, $r<0$, when the system is a metal, the dc conductance at zero temperature does not contain any information about the universal signatures of the Mott transition because it is solely determined by the compressibility of the electron liquid in the leads, i.e., by the Luttinger parameter $K_L$  in agreement with Refs.~\onlinecite{Starykh98} and \onlinecite{Mori97}. However, in the insulating phase, $r>0$, we found evidence that the 
critical degress of freedom of the quantum phase transition become manifest in 
the dc transport. In particular, in the lowest order in the fugacity $e^{-r/T}$, the conductance is independent of the  properties of the leads, $K_L$, while the exponential prefactor is determined by the universal characteristic of the Mott transition as it is given by an effective conductance quantum of the fractional charge carriers, $G_{\rm CI} = 2 e^2_{\rm CI}/h$. The conductance in this regime originates from thermal activation of solitonic excitations that travel ballistically across the Mott-gapped region of the wire. 
Finally, in the quantum critical regime, $|r| < T$, of the phase diagram in Fig.~\ref{fig:1} the conductance is dominated by its critical value
(\ref{Gcr}) that depends on the Luttinger parameter of the leads, $K_L$, 
as well as on the fractional charge $e_{\rm CI}$. 

Our main finding is that the dc transport shows signatures of charge fractionalization (\ref{FracCharge}) in the Mott wire. The underlying reason is that the electrons, irrespective of their interactions in the leads, must ``transform'' into Mott solitons with charge (\ref{FracCharge}) in order to overcome the Mott barrier $\Delta$. 
The fractional charge can in principle be detected with the help of the finite temperature behavior of the critical two-terminal conductance of a Mott wire. 
We contrast this fractionalization with that in a Luttinger-liquid wire, where 
only the forward scattering is included;\cite{FG} there, the phenomenon of fractionalization is not manifest in the dc conductance.\cite{Safi95,Maslov95,Ponomarenko95}

The regime discussed above is restricted to the immediate vicinity of the quantum critical point of Fig.~\ref{fig:1}, i.e. to the temperatures and to the tuning parameter much smaller than the Mott gap: $|r|,T \ll \Delta$. In particular, we neglected corrections arising from the thermal activation of particles in the other Mott band which are exponentially small in $e^{-(2 \Delta-r)/T}$. Furthermore, we disregarded effects of tunneling of solitons through the Mott barrier, which gives rise to a finite conductance in the insulating regime of the form $\delta G \sim e^{-S}$, where $S \propto r L/v$ with the length of the barrier $L$ and the velocity of solitons $v$. The approximation of neglecting this tunneling correction breaks down at sufficiently low temperatures when the length $L$ becomes comparable to the thermal wavelength $\xi_T = v/T$.

We arrived at our conclusions by performing a controlled perturbative
expansion around the special value $K_L=1/2$ of the Luttinger
parameter in the leads. At this specific Luther-Emery value, the
solitonic excitations are free, and the problem of the Mott wire with
leads reduces to a 1D scattering problem as descibed in Section
\ref{sec:Leads1/2}.  For $K_L$ different from the Luther-Emery value,
the interaction between the scattering states in the leads yields the
result (\ref{FinalResult}).
   
Our conclusions concerning the temperature dependence of the critical conductance differ from the results of the work of Mori {\it et al.}.\cite{Mori97} There, a similar setup was considered, and the current autocorrelation function was also calculated in terms of the critical degrees of freedom of the Mott transition. However,
the current correlator was only considered for a homogeneous system when the Mott gap extends across the full wire length resulting in a translationally invariant expression. 
The result was then used to match the current autocorrelator within the Mott wire with the one of the Fermi liquid leads in order to derive the conductance. 
We believe, however, that the use of the current autocorrelator for the homogeneous Mott wire is inconsistent with the matching conditions applied in Ref.~\onlinecite{Mori97}, which explicitly break translationary invariance.
Instead of using the homogeneous current autocorrelation function 
one may consider the result (\ref{Polarizations}) 
for the case of a gapped region inside the wire as a starting point 
for a matching procedure. Preliminary considerations along these lines have led us to the conjecture described below.

Let us speculate about the functional form of dc conductance beyond the leading $K_L -\frac{1}{2}$ correction.  
We conjecture that, 
in the limit $|r|, T \ll \Delta$ and $T \gg v/L$,
the corresponding resistance is a sum of the two contributions
\begin{align} \label{Conjecture}
R &= R_L + R_{M}\quad {\rm with}
\\ \non
R_L &= \frac{1}{K_L} \frac{h}{2 e^2},\quad{\rm and}\quad
R_M = \frac{h}{2 e^2_{\rm CI}} \frac{1-f(r/T)}{f(r/T)} \,,
\end{align}
where $e_{\rm CI} = e/\sqrt{2}$ and 
$f$ is the Fermi function (\ref{FermiFunction}).
The resistance $R_L$ is attributed to the leads and can be interpreted as a contact resistance. In particular, only $R_L$ depends on the Luttinger parameter within the leads, $K_L$. 
The contribution $R_M$ has the form of a four-terminal resistance with an effective transmission given by the Fermi function, $f(r/T)$. 
It originates from the critical Mott part of the wire and involves the charge quantum of the critical degrees of freedom, $e_{\rm CI} $. The expression (\ref{Conjecture}) reproduces the known limits. In the incommensurate phase, $r<0$, at $T=0$ the resistance reduces to the contact resistance $R =  R_L$. In the insulating state, $r>0$, at lowest temperature the resistance is dominated by the contribution of the Mott region $R_M$ and becomes independent of the lead properties. Expanding (\ref{Conjecture}) in $K_L - \frac{1}{2}$ one also reproduces our hard-to-guess perturbative result (\ref{FinalResult}).

The Mott region in a quantum wire may be realized by employing an external
periodic potential created by gating,\cite{kouwenhoven} 
or an acoustic field.\cite{SAW}
  
Our approach applies also to the problem of 
drag between two quantum wires.
At equal electron densities, the inter-wire interaction
locks the charges in both wires such that the drag resistance at $T=0$
is infinite in the commensurate state. 
This system bears resemblance to the Mott insulator: if opposite biases are 
applied to the two wires, there is no current at $T=0$.\cite{Averin-Nazarov,Klesse-Stern} Density imbalance drives the system into an incommensurate phase, for which
one recovers the ballistic conductance for each of the wires; drag is absent at 
$T=0$.
The problem of a finite-$T$ drag resistance
may be mapped onto that of the critical conductance in a Mott wire. Technically,
such a mapping is established by introducing the even and odd modes, 
$\rho_\pm = (\rho_1\pm \rho_2)/\sqrt{2}$, 
where $\rho_{1,2}(x)$ are the electron densities
in the two wires. 
The dynamics of the odd mode is described by the 
sine-Gordon model (\ref{SineGordon}), while the even mode is a conventional
Luttinger liquid. The drag resistance is determined by
the conductance in the odd sector. The present analysis of 
the critical behavior of the transport
at the commensurate-incommensurate transition is fully applicable for the 
evaluation of the drag resistance.

\acknowledgments

M.G. was supported by SFB 608 of the DFG. 
D.N. and L.G. were supported by NSF grants 
DMR 02-37296 and DMR 04-39026. 
A.S. acknowledges financial support by the Israel Science Foundation, 
administered by the Israeli Academy of Sciences.

\appendix

\section{Polarization operator in the coordinate-frequency representation}
\label{app:pol}

We present a detailed calculation of the polarization operators (\ref{Pi-ij-def}).
After performing the sum over fermionic Matsubara frequencies we obtain 
\be \label{Pi-ij}
\Pi_{x,x';\W_n}^{ij} = \int d\epsilon d\epsilon'
{f(\epsilon)-f(\epsilon')\over i\W_n - (\epsilon-\epsilon')}
\tr \lf \tau^i \mathbf{A}_{x,x';\epsilon}\tau^j \mathbf{A}_{x',x;\epsilon'} \rf
\ee
with the spectral function $\mathbf{A}$ given in (\ref{A}). In this section, we use a temperature dependent definition of the Fermi function, $f(\omega) = 1/(e^{\omega/T}+1)$. As a first step we evaluate the trace. We will use the abbreviations (\ref{ss'}).
Consider first a situation with $s \neq s'$, i.e.~with $x$ and $x'$ on opposite sides of the barrier. 
\begin{align}
s\neq s':\quad&
\tr \lf \tau^i \mathbf{A}_{x,x';\epsilon}\tau^j \mathbf{A}_{x',x;\epsilon'} \rf 
\\\non
&= 
\nu^2 \Theta_{\epsilon-r}\Theta_{\epsilon'-r}  
\A^{ij}_{\epsilon,\epsilon'}(x-x' +s \ell_t) \,.
\label{tr:s<>s'}
\end{align}
Here
\be \label{Aij}
\A^{ij}_{\epsilon,\epsilon'}(x) = 
e^{i(\epsilon-\epsilon')x/v} + \beta^i\beta^j e^{-i(\epsilon-\epsilon')x/v} \,,
\ee
where the $\beta^i$ were defined in (\ref{SigmaMatrices}); and we made use of the approximations (\ref{Phases}) for the phase shifts with $\ell_t = (\ell_+ + \ell_-)/2$. The case $s=s'$ is done explicitly 
[in Eq.~(\ref{tr:s=s'}) 
set $v=1$ for brevity]:
\begin{align}
\label{tr:s=s'}
\lefteqn{s=s': \quad
\tr \lf \tau^i \mathbf{A}_{x,x';\epsilon}\tau^j \mathbf{A}_{x',x;\epsilon'} \rf =}
\\ &= \nu^2 \tr 
\begin{bmatrix} 
e^{i\epsilon(x-x')} & s\Theta_{r-\epsilon}e^{i\epsilon(x+x')+i s \phi_s(\epsilon)}
\\
s\beta^i \Theta_{r-\epsilon}e^{-i\epsilon(x+x')-i s \phi_s(\epsilon)} 
& \beta^i e^{-i\epsilon(x-x')} 
\end{bmatrix} 
\non
\\ &\times
\non
\begin{bmatrix} 
e^{-i\epsilon'(x-x')} & 
s\Theta_{r-\epsilon'}e^{i\epsilon'(x+x')+i s \phi_s(\epsilon')} \\
s\beta^j \Theta_{r-\epsilon'}e^{-i\epsilon'(x+x')-i s \phi_s(\epsilon')} 
& \beta^j e^{i\epsilon'(x-x')} 
\end{bmatrix} 
 \\ \non
&=\nu^2\left(\A_{\epsilon,\epsilon'}^{ij}(x-x') 
+ \beta^j \Theta_{r-\epsilon}\Theta_{r-\epsilon'}
\A_{\epsilon,\epsilon'}^{ij}(x+x'+ s \ell_s) \right) \quad\quad
\end{align}
where we used the approximation (\ref{Phases}) 
for the reflection phase shifts.

When substituting the above expressions for the trace 
into Eq.~(\ref{Pi-ij}), it is convenient to define 
\be \label{F-Delta}
F_r(x) = \nu^2 \int_{r}^\infty \! d\epsilon d\epsilon'\, 
{f(\epsilon)-f(\epsilon')\over i\W_n - (\epsilon-\epsilon')} 
e^{i(\epsilon-\epsilon')x/v} \,.
\ee
Note that in both cases, {\it after} evaluating the trace,
the conditions $\epsilon>r$ and $\epsilon'>r$,
or  $\epsilon<r$ and $\epsilon'<r$, enter simultaneously.
Hence the polarization operator (\ref{Pi-ij}) can be expressed via (\ref{F-Delta})
as 
\begin{align} \label{Pi-F}
\lefteqn{\Pi_{x,x';\W_n}^{ij} =}
\\ \non 
= & {1-ss'\over 2} \lb F_r(x-x'+s \ell_t) + \beta^i\beta^j F_r(x'-x-s \ell_t)\rb
\\ \non
 &+ {1+ss'\over 2} \lf \Pi^{(0)ij}_{x,x';\Omega_n} 
\right. 
\\ \non 
&+ \left.\beta^j  \lb F_{-r}(x+x' + s \ell_s) + \beta^i\beta^j F_{-r}(-x'-x- s \ell_s)\rb 
\vphantom{\Pi_0^{ij}(x)}\!\! \rf .
\quad\quad
\end{align}
The last term is obtained from Eq.~(\ref{F-Delta}) via the interchange
of the dummy variables $\epsilon\to -\epsilon'$ and $\epsilon'\to -\epsilon$,
and using $f(-\epsilon)=1-f(\epsilon)$.
We also define the polarization operator in the translation-invariant case
\be \label{Pi0-ij}
 \Pi^{(0)ij}_{x,x';\Omega_n}  = \lb F_{r}(x-x') 
+ \beta^i\beta^j F_r(x'-x)\rb_{r\to -\infty} \,.
\ee

We now proceed to evaluating (\ref{F-Delta}). Introducing the 
variable $\varepsilon=\epsilon-\epsilon'$, write
\bea  \label{FOrderOfIntegrations}
F_r(x) &=& 
\nu^2 \int_{r}^\infty \! d\epsilon 
\int_{-\infty}^{\epsilon-r} d\varepsilon \, 
{f(\epsilon)-f(\epsilon-\varepsilon)\over i\W_n - \varepsilon} 
e^{i \varepsilon x/v} 
\\ \non
&=& \nu^2 \int_{-\infty}^0 {e^{i \varepsilon x/v}  d\varepsilon \over i\W_n - \varepsilon} 
\int_{r}^\infty \! d\epsilon \lb f(\epsilon)-f(\epsilon-\varepsilon)\rb 
\\ \non
&+& \nu^2 \int_0^{\infty} {e^{i \varepsilon x/v} d\varepsilon \over i\W_n - \varepsilon} 
\int_{r+\varepsilon}^\infty \! d\epsilon \lb f(\epsilon)-f(\epsilon-\varepsilon)\rb .
\eea
The integration with respect to  $\epsilon$ and $\varepsilon$ is over a domain that can be split into an infinite rectangle, $\epsilon > r$ and $\varepsilon < 0$, and an infinite triangle, $\epsilon > r$ and $0 < \varepsilon < \epsilon -r$. After changing the order of integration in each of the two sub-domains separately, such that the integral over $\epsilon$ is performed first, one obtains the last equation in (\ref{FOrderOfIntegrations}). 
Shifting variables in the energy integrations and introducing 
the function
\be \label{varphi}
\varphi(\varepsilon) = \int_r^{r+\varepsilon}\! d\epsilon\, f(\epsilon) 
= \varepsilon + T \ln {f(r+\varepsilon)\over f(r)} \,,
\ee
we get
\begin{align}
F_r(x) = \nu^2 \int_0^\infty d\varepsilon\,  \varphi(\varepsilon) 
\lb {e^{i \varepsilon x/v}  \over \varepsilon-i\W_n} + {e^{-i \varepsilon x/v}  \over \varepsilon+ i\W_n} \rb.
\end{align}
Consider first the static part, $F^{\rm stat}_r(x) \equiv \left.F_r(x)\right|_{i \Omega_n = 0}$; it can be represented as 
\begin{align} \label{staticPart}
F^{\rm stat}_r(x) = \nu f(r) \delta_B(x)
\end{align}
where $\delta_B(x)$ is a function with a peak at $x=0$ that carries unit weight, 
$\int \! dx\, \delta_B(x) \equiv 1$. In the limit $r\ \to -\infty$, the function $\delta_B$ reduces to a delta function; in the classical limit, $0 < r/T \ll 1$, the peak has a width of order $\xi_T = v/T$.

The dynamic part of the $F_r$ function, $F^{\rm dyn}_r \equiv F_r - F^{\rm stat}_r$,
\begin{align}
F^{\rm dyn}_r(x) = \nu^2 i \Omega_n \int_0^\infty d\varepsilon\,  \frac{\varphi(\varepsilon)}{\varepsilon} 
\lb {e^{i \varepsilon x/v}  \over \varepsilon-i\W_n} - {e^{-i \varepsilon x/v}  \over \varepsilon+ i\W_n} \rb
\end{align}
is dominated by the poles, $\varepsilon = \pm i \Omega_n$. 
The leading behavior
%
\begin{align} \label{dynamicPart}
F^{\rm dyn}_r(x) \simeq - \pi \nu^2 \W_n f(r)  
\lb \sign x + \sign \W_n\rb
e^{-|\W_n x|/v}
\end{align}
originates from the odd part
$\varphi(\varepsilon)\to \varphi_-=\frac12 [\varphi(\varepsilon)-\varphi(-\varepsilon)]
= \varepsilon f(r) + {\cal O}(\varepsilon^3)$
by extending the integral over to the real axis and 
evaluating the Taylor-expanded $\varphi_-(\varepsilon)$.
The even part, $\varphi_+=\frac12 [\varphi(\varepsilon)+\varphi(-\varepsilon)]$,
and the higher-order terms of the odd part, yield the corrections 
$\mathcal{O}(f'(r) \Omega_n^2 \log |\Omega_n x|)$ that are subleading 
in the dc limit, $\omega/T \ll 1$,
of small real frequencies, $i\Omega_n \to \omega + i 0$ 
after the analytic continuation. 

We now have all the ingredients to construct the polarization operator
(\ref{Pi-F}). First, the absolute values of the arguments of the $F_r$ functions in (\ref{Pi-F}) are all larger than the scattering lengths $\ell_\pm$ that characterize the length of the Mott region of the wire. In the limit of a long Mott wire, $\ell_\pm T/v \gg 1$ we can neglect the tails of the static part (\ref{staticPart}) of the $F_r$ functions in (\ref{Pi-F}). The static part then only enters the polarization via the contribution $\Pi^{(0)ij}$ which was defined in (\ref{Pi0-ij}). The $F_r$ functions in (\ref{Pi-F}) thus only contribute with their dynamic part (\ref{dynamicPart}). Moreover, for the dynamic part the scattering lengths $\ell_\pm$ in the arguments of the $F_r$ functions in (\ref{Pi-F}) can be effectively set to zero as they give only subleading corrections in the dc limit $\omega \ell_\pm/v \ll 1$. After these simplifications we finally obtain the expression (\ref{Pi-Pi0}) for the polarization operator.



\begin{thebibliography}{99}
\bibitem{footnote-gap}
The bare gap $\Delta$ in the Hamiltonian (\ref{MT}) is renormalized
by the interactions. The details of such renormalization are not important;
here we assume that the tuning parameter (\ref{-r})
is defined with respect to the renormalized $\Delta$.
\end{thebibliography}
\end{document}